\documentclass{article}

\usepackage{amsmath,amsthm,amssymb}

\usepackage[dvips]{graphicx}

\def\R{\mathbb R}
\def\N{\mathbb N}
\def\C{\mathbb C}
\def\Z{\mathbb Z}

\def\pf{\begin{proof}}
\def\pfk{\end{proof}}

\newtheorem{lem}{Lemma}[section]
\newtheorem{prop}[lem]{Proposition}
\newtheorem{thm}[lem]{Theorem}
\newtheorem{coro}[lem]{Corollary}

\theoremstyle{definition}

\newtheorem{pozn}[lem]{Remark}

\renewcommand{\i}{{\mathrm i}}
\def\e{\mathrm e}

\def\O{\mathcal O}
\def\G{\mathcal{G}}

\begin{document}

\begin{center}
{\Large\bf On the spectrum of a bent chain graph}

\vspace{0.2cm}
\bigskip
{\large Pierre Duclos$^1$, Pavel Exner$^{2,3}$, Ond\v{r}ej
Turek$^4$}

\bigskip
{1) Centre de Physique Th\'eorique de Marseille UMR
  6207 - Unit\'e Mixte de Recherche du CNRS et des Universit\'es
  Aix-Marseille I, Aix-Marseille II et de l' Universit\'e du Sud
  Toulon-Var - Laboratoire affili\'e \`a la FRUMAM\\
2) Doppler Institute, Czech Technical University, B\v{r}ehov\'{a}
7, 11519 Prague, Czech Republic \\
3) Department of Theoretical Physics, NPI, Czech Academy of
Sciences, 25068~\v{R}e\v{z} near Prague, Czech Republic \\
4) Department of Mathematics, FNSPE, Czech Technical University,
Trojanova 13, 12000 Prague, Czech Republic\\
\emph{duclos@univ-tln.fr, exner@ujf.cas.cz,
turekond@fjfi.cvut.cz}}
\end{center}

\bigskip
\begin{abstract}
We study Schr\"odinger operators on an infinite quantum graph of a
chain form which consists of identical rings connected at the
touching points by $\delta$-couplings with a parameter
$\alpha\in\R$. If the graph is ``straight'', i.e. periodic with
respect to ring shifts, its Hamiltonian has a band spectrum with
all the gaps open whenever $\alpha\ne 0$. We consider a
``bending'' deformation of the chain consisting of changing one
position at a single ring and show that it gives rise to
eigenvalues in the open spectral gaps. We analyze dependence of
these eigenvalues on the coupling $\alpha$ and the ``bending
angle'' as well as resonances of the system coming from the
bending. We also discuss the behaviour of the eigenvalues and
resonances at the edges of the spectral bands.
\end{abstract}

\section{Introduction}

Quantum graphs, i.e. Schr\"odinger operators with graph
configuration spaces, were introduced in the middle of the last
century \cite{RuS53} and rediscovered three decades later
\cite{GP, ES}. Since then they attracted a lot of attention; they
became both a useful tool in numerous applications and a mean
which makes easy to study fundamental properties such as quantum
chaos. We refrain from giving an extensive bibliography and refer
to the recent proceedings volume \cite{AGA} which the reader can
use to check the state of art in this area.

One of the frequent questions concerns relations between the
geometry of a graph $\Gamma$ and spectral properties of a
Schr\"odinger operator supported by $\Gamma$. Put like that, the
question is a bit vague and allows different interpretation. On
one hand, we can have in mind the intrinsic geometry of $\Gamma$
which enters the problem through the adjacency matrix of the graph
and the lengths of its edges. On the other hand, quite often one
thinks of $\Gamma$ as of a subset of $\mathbb{R}^n$ with the
geometry inherited from the ambient space. In that case geometric
perturbations can acquire a rather illustrative meaning and one
can ask in which way they influence spectral properties of a
quantum particle ``living'' on $\Gamma$; in such a context one can
think of graphs with various local deformations as ``bent'', locally ``protruded'' or ``squeezed'', etc.

This is particularly interesting if the ``unperturbed'' system is
explicitly solvable being, for instance, an infinite periodic
graph. An influence of local spectral perturbations mentioned
above is in this setting a rich subject which deserves to be
investigated. So far it was considered only episodically but even
such a brief look shows that it may have properties uncommon in
the usual theory of Schr\"odinger operators \cite{KV}. With this
motivation we find it useful to start such a programme by
discussing the influence of a ``bending'' deformation on a graph
which exhibits a one-dimensional periodicity.

\begin{figure}[h]
\vspace{2em}
\begin{center}
\includegraphics[angle=0,width=0.7\textwidth]{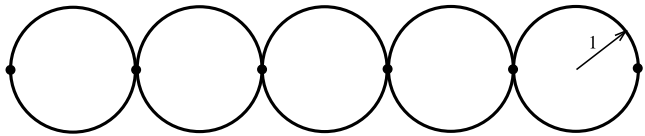} \vspace{-4em}
\caption{The unperturbed chain graph}\label{puvodni}
\end{center}
\end{figure}

To make things as simple as possible at the beginning we will not
strive in this paper for generality and we will discuss in detail
a simple nontrivial example, allowing for a fully explicit
solution, in which the unperturbed system is a ``chain graph''
consisting of an array of rings of unit radius,
cf.~Fig.~\ref{puvodni}, connected through their touching points.
We suppose that there are no external fields. Since values of
physical constants are not important in our considerations we put
$\hbar=2m=1$ and identify the particle Hamiltonian with the
(negative) Laplacian acting as $\psi_j\mapsto -\psi_j''$ on each
edge of the graph. It is well known that in order to get a
self-adjoint operator one has to impose appropriate boundary
conditions at the graph vertices. In our model we employ the
so-called $\delta$-\emph{coupling} characterized by the conditions
\begin{equation} \label{delta}
\psi_j(0)=\psi_k(0)=:\psi(0)\,, \quad j,k\in\hat{n}\,,
\qquad \sum^{n}_{j=1}\psi_j'(0)=\alpha\psi(0)\,,
\end{equation}
where $\hat{n}=\{1,2,\ldots,n\}$ is the index set numbering the
edges emanating from the vertex --- in our case $n=4$ --- and
$\alpha\in\R\cup\{+\infty\}$ is the coupling constant supposed to
be the same at every vertex of the chain. It is important that the
``straight'' graph has spectral gaps\footnote{A nontrivial vertex
coupling is also related to the problem of approximation of
quantum graphs by ``fat graphs'' of which the reader can learn
more, e.g., in \cite{CE} or \cite{EP}, references therein, and a paper in preparation by the last mentioned authors.}, thus we \emph{exclude the free boundary conditions} (sometimes called, not quite appropriately,
Kirchhoff), i.e. we assume $\alpha\ne0$.

\begin{figure}[h]
\begin{center}
\includegraphics[angle=0,width=0.9\textwidth]{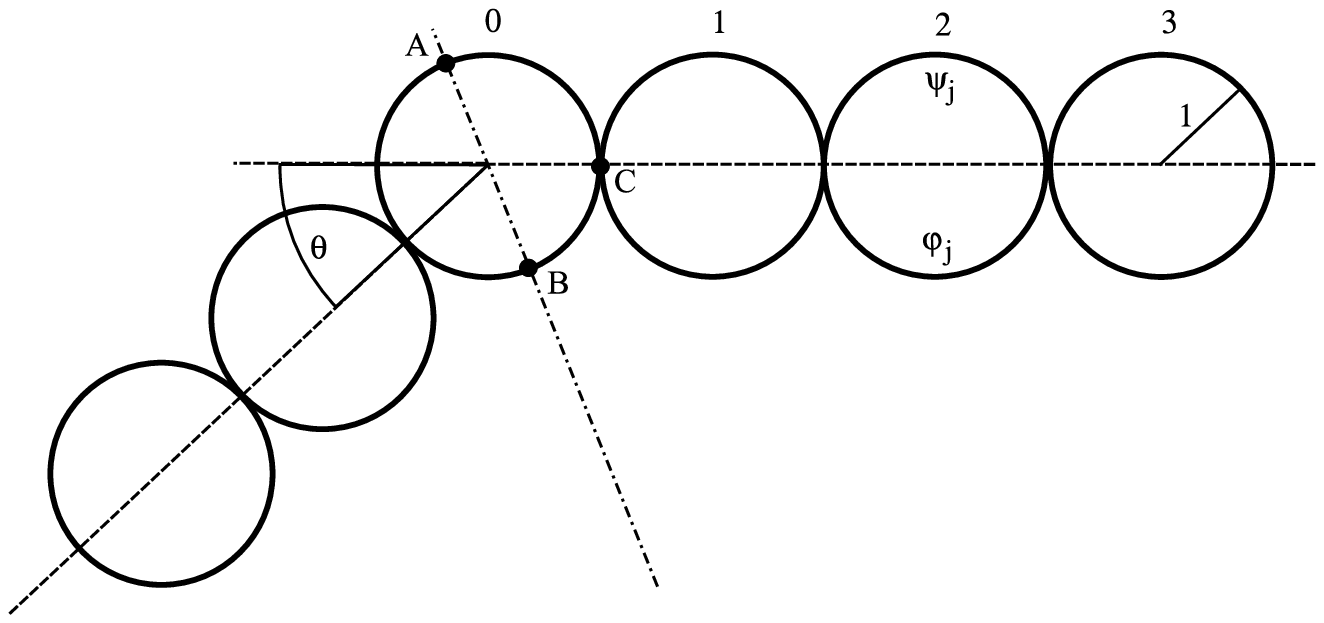}
\caption{A bent graph}\label{ohyb}
\end{center}
\end{figure}

The geometric perturbation to consider is the simplest possible
bending of such a chain obtained by a shift of one of the contact
points, as sketched in Fig.~\ref{ohyb}, which is parametrized by
the bending angle $\vartheta$ characterizing the ratio of the two
edges constituting the perturbed ring.
Our aim is to show that the bending gives rise to eigenvalues in
the gaps of the unperturbed spectrum and to analyze how they
depend on $\vartheta$. At the same time the bent chain will
exhibit resonances and we will discuss behaviour of the
corresponding poles.

The contents of the paper is the following. In the next section we
analyze the straight chain. Using Bloch-Floquet decomposition we
will show that the spectrum consists of infinite number of
absolutely continuous spectral bands separated by open gaps, plus
a family of infinitely degenerate eigenvalues at band edges. In
Section~\ref{perturbace} we will analyze the discrete spectrum due
to the bending showing that in each gap it gives rise to at most
two eigenvalues. Section~\ref{reson} describes their dependence on
the bending angle as well as complex solutions to the spectral
condition corresponding to resonances in the bent chain. In
Section~\ref{angle} we discuss further the angular dependence with
attention to singular points where the solutions coincide with the
band edges. Finally, in the concluding remarks we draw a parallel
of our results with properties of quantum waveguides.

\setcounter{equation}{0}
\section{An infinite periodic chain}

First we consider a ``straight'' chain $\Gamma_0$ as sketched in
Fig.~\ref{puvodni}; without loss of generality we may suppose that
the circumference of each ring is $2\pi$. The state Hilbert space
of a nonrelativistic and spinless particle living on $\Gamma_0$ is
$L^2(\Gamma_0)$. We suppose that the particle is free, not
interacting with an external potentials on the edges, and denote
by $H_0$ its Hamiltonian, i.e. it acts as the negative Laplacian
on each graph link and its domain consists of all functions from
$W^{2,2}_\mathrm{loc}(\Gamma_0)$ which satisfy the $\delta$
boundary conditions (\ref{delta}) at the vertices of $\Gamma_0$;
we suppose that the coupling constant $\alpha$ is the same at each
vertex\footnote{The coupling constant $\alpha$ is kept fixed and
for the sake of simplicity we will not use it to label the
Hamiltonian neither in the straight nor in the bent case.}.

\begin{figure}[h]
\begin{center}
\includegraphics[angle=0,width=0.2\textwidth]{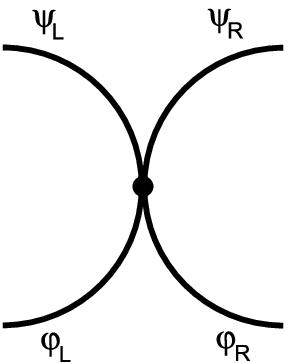}
\caption{Elementary cell of the periodic system}\label{bunka}
\end{center}
\end{figure}

In view of the periodicity of $\Gamma_0$, the spectrum of $H_0$
can be computed using Bloch-Floquet decomposition. Let us consider
an elementary cell with the wavefunction components denoted
according to the Fig.~\ref{bunka} and ask about the spectrum of
the Floquet components of $H_0$. Since the operator acts as a
negative second derivative, each component of the eigenfunction
with energy $E=k^2\neq0$ is a linear combination of the functions
$\e^{\pm\i k x}$. The momentum $k$ is conventionally chosen
positive for $E>0$, while for $E$ negative we put $k=\i\kappa$
with $\kappa>0$ (the case $E=0$ will be mentioned separately
below). For a given $E\neq0$, the wavefunction components on the
elementary cell are therefore given by
\begin{equation}\label{vlnfce}
\begin{split}
\psi_L(x)&=C_L^+\e^{\i k x}+C_L^-\e^{-\i k x},\quad x\in[-\pi/2,0]\\
\psi_R(x)&=C_R^+\e^{\i k x}+C_R^-\e^{-\i k x},\quad x\in[0,\pi/2]\\
\varphi_L(x)&=D_L^+\e^{\i k x}+D_L^-\e^{-\i k x},\quad x\in[-\pi/2,0]\\
\varphi_R(x)&=D_R^+\e^{\i k x}+D_R^-\e^{-\i k x},\quad x\in[0,\pi/2]
\end{split}
\end{equation}
As we have said, at the contact point the $\delta$-coupling
(\ref{delta}) is assumed, i.e.
\begin{gather}
\psi_L(0)=\psi_R(0)=\varphi_L(0)=\varphi_R(0) \nonumber \\[-.6em]
\label{vazba} \\[-.6em]
-\psi_L'(0)+\psi_R'(0)-\varphi_L'(0)+\varphi_R'(0)=\alpha\cdot\psi_L(0)
\nonumber
\end{gather}
On the other hand, at the ``free'' ends of the cell the Floquet
conditions are imposed,
\begin{equation}\label{Floquet}
\begin{split}
\psi_R(\pi/2)=\e^{\i \theta}\psi_L(-\pi/2)\,&\qquad
\psi_R'(\pi/2)=\e^{\i \theta}\psi_L'(-\pi/2)\\
\varphi_R(\pi/2)=\e^{\i \theta}\varphi_L(-\pi/2)\,&\qquad
\varphi_R'(\pi/2)=\e^{\i \theta}\varphi_L'(-\pi/2)
\end{split}
\end{equation}
with $\theta$ running through $[-\pi,\pi)$; alternatively we may
say that the quasimomentum $\frac{1}{2\pi}\theta$ runs through
$[-1/2,1/2)$, the Brillouin zone of the problem.

Substituting~\eqref{vlnfce} into~\eqref{vazba} and
\eqref{Floquet}, one obtains after simple manipulations
\begin{equation}\label{shodnost}
C_X^+\cdot\sin k\pi=D_X^+\cdot\sin k\pi\,,\quad C_X^
-\cdot\sin k\pi=D_X^-\cdot\sin k\pi\,,
\end{equation}
where $X$ stands for $L$ or $R$, hence $C_X^+=C_X^-$ and
$D_X^+=D_X^-$ provided $k\notin\N_0:=\{0,1,2,\dots\}$. We will
treat the special case $k\in\N$ later, now we will suppose $k$
does not belong to $\N$, the set of natural numbers. Furthermore,
from~\eqref{vazba} and~\eqref{Floquet} we obtain an equation for
the phase factor $\e^{\i \theta}$,
\begin{equation}\label{eik}
\e^{2\i \theta}-\e^{\i \theta}\left(2\cos k\pi
+\frac{\alpha}{2k}\sin k\pi\right)+1=0\,,
\end{equation}
which has real coefficients for any
$k\in\R\cup\i\R\backslash\{0\}$ and the discriminant equal to
$$
D=\left(2\cos k\pi+\frac{\alpha}{2k}\sin k\pi\right)^2-4\,.
$$
We have to determine values of $k^2$ for which there is a
$\theta\in[-\pi,\pi)$ such that~\eqref{eik} is satisfied, in other
words, for which $k^2$ it has, as an equation in the unknown
$\e^{\i \theta}$, at least one root of modulus one. Note that a
pair of solutions of~\eqref{eik} always give one when multiplied,
regardless the value of $k$, hence either both roots are complex
conjugated of modulus one, or one is of modulus greater than one
and the other has modulus smaller than one. Obviously, the latter
situation corresponds to a positive discriminant, and the former
one to the discriminant less or equal to zero. We summarize this
discussion as follows:
\begin{prop}\label{pasy}
If $k^2\in\R\backslash\{0\}$ and $k\notin\N$, then
$k^2\in\sigma(H_0)$ if and only if the condition
\begin{equation}\label{sporig}
\left|\cos k\pi+\frac{\alpha}{4}\cdot\frac{\sin k\pi}{k}\right|\leq1
\end{equation}
is satisfied.
\end{prop}

In particular, the negative spectrum is obtained by putting
$k=\i\kappa$ for $\kappa>0$ and rewriting the
inequality~\eqref{sporig} in terms of this variable. Note that
since $\sinh x\neq0$ for all $x>0$, it never occurs that $\sin
k\pi=0$ for $k\in\i\R^+$, the positive imaginary axis, thus there
is no need to treat this case separately like for $k\in\R^+$,
cf.~\eqref{shodnost} above.

\begin{coro}
If $\kappa>0$, then $-\kappa^2\in\sigma(H_0)$ if and only if
\begin{equation}\label{zaporne}
\left|\cosh\kappa\pi+\frac{\alpha}{4}\cdot
\frac{\sinh\kappa\pi}{\kappa}\right|\leq1\,.
\end{equation}
\end{coro}

Let us finally mention the case $k\in\N$ left out above. It is
straightforward to check that $k^2$ is then an eigenvalue, and
moreover, that it has an infinite multiplicity. One can construct
an eigenfunction which is supported by a single circle, which is
given by $\psi(x)=\sin k x$ with $x\in[0,\pi]$ on the upper
semicircle and $\varphi(x)=-\sin k x$ with $x\in[0,\pi]$ on the
lower one.

\begin{pozn} \label{updown}
The condition~\eqref{sporig} reminds us of the corresponding
condition in the Kronig-Penney model with the distance between the
interaction sites equal to $\pi$, cf.~\cite{AGHH}, the only
difference being that the coupling constant is halved,
$\frac12\alpha$ instead of $\alpha$. In contrast to that, the
point spectrum of the KP model is empty. These facts are easy to
understand if we realize that our model has the up-down mirror
symmetry, and thus $H_0$ decomposes into a symmetric and
antisymmetric part. The former is unitarily equivalent to the KP
model with modified coupling, the latter corresponds to functions
vanishing at the vertices, having thus a pure point spectrum.
Looking ahead, we remark that the bending perturbation breaks this
mirror symmetry.
\end{pozn}

\noindent Finally, in the case $E=0$ we get in the similar way the
equation
\begin{equation}\label{ei0}
\e^{2\i \theta}-\e^{\i
\theta}\left(2+\frac{\alpha\pi}{2}\right)+1=0\,,
\end{equation}
replacing~\eqref{eik}, whence we infer that $0\in\sigma(H_0)$ if
and only if
\begin{equation}\label{alphanula}
\left|1+\frac{\alpha\pi}{4}\right|\leq1\,,
\end{equation}
hence zero can belong to the continuous part of the spectrum only
and it happens \emph{iff} $\alpha\in [-8/\pi,0]$. In conclusion,
we can make the following claim about $\sigma(H_0)$.

\begin{thm}\label{bands}

The spectrum of $H_0$ consists of infinitely degenerate
eigenvalues equal to $n^2$ with $n\in\N$, and absolutely
continuous spectral bands with the following properties: \\[.5em]
If $\alpha>0$, then every spectral band is contained in an
interval $(n^2,(n+1)^2]$ with $n\in\N_0:=\N\cup\{0\}$, and its
upper edge
coincides with the value $(n+1)^2$. \\[.5em]
If $\alpha<0$, then in each interval $[n^2,(n+1)^2)$ with $n\in\N$
there is exactly one spectral band the lower edge of which
coincides with $n^2$. In addition, there is a spectral band with
the lower edge (being the overall spectral threshold) equal to
$-\kappa^2$, where $\kappa$ is the largest solution of
 \begin{equation} \label{lowband}
\left|\cosh\kappa\pi+\frac{\alpha}{4}
\cdot\frac{\sinh\kappa\pi}{\kappa}\right|=1\,.
 \end{equation}
The position of the upper edge of this band depends on $\alpha$.
If $-8/\pi<\alpha<0$, then it is equal to $k^2$ where $k$ is the
solution of
$$
\cos k\pi+\frac{\alpha}{4}\cdot\frac{\sin k\pi}{k}=-1
$$
contained in $(0,1)$. On the other hand, for $\alpha<-8/\pi$ the
upper edge is negative, $-\kappa^2$ with $\kappa$ being the
smallest solution of \eqref{lowband}, and for $\alpha=-8/\pi$ it
equals zero. \\[.5em]
Finally, $\sigma(H_0)=[0,+\infty)$ holds if $\alpha=0$.

\end{thm}
\pf The degenerate bands, in other words, the eigenvalues of
infinite multiplicity, were found already and it is
straightforward to check that no other eigenvalues exist. The
continuous spectrum can be in view of Remark~\ref{updown} treated
as in \cite{AGHH}, nevertheless, we sketch the argument not only
to make the paper self-contained, but also in view of next
sections where some ideas and formula of the present proof will be
used again.

Consider first the positive part of the continuous spectrum. The
condition~\eqref{sporig} clearly determines bands with one
endpoint at $n^2,\: n\in\N$, where the sign of $\alpha$ decides
whether it is the upper or lower one. If $\alpha<0$, the presence
of a band in (0,1) depends on $|\alpha|$. Denoting $g(x):= \cos
x\pi+\frac{\alpha}{4}\cdot\frac{\sin x\pi}{x}$ we want to show
that $B:=\{x\in(0,1):\: |g(x)|\leq1\}$ is either empty or an
interval with zero as its edge. It is obvious that $g$ maps
$(0,1)$ continuously into $(-\infty,1)$; we will check that
$g(x_0)=-1$ implies $g'(x_0)<0$. We notice first that the premise
implies $\cos x_0\pi= -1-\frac{\alpha}{4}\cdot\frac{\sin
x_0\pi}{x_0}$; taking the square of this relation we find after
simple manipulations that $\sin x_0\pi=
-2\left(\frac{\alpha}{4x_0} +\frac{4x_0}{\alpha}\right)^{-1}$ and
$\cos x_0\pi= \left(\frac{\alpha}{4x_0}-\frac{4x_0}{\alpha}\right)
\left(\frac{\alpha}{4x_0}+\frac{4x_0}{\alpha}\right)^{-1}$.
Evaluating $g'(x_0)$ and substituting these expressions we get
$$
g'(x_0)=\frac{\alpha\pi}{4x_0}\left(1-\frac{\sin\pi x_0}{\pi
x_0}\right)<0\,.
$$
These properties together with the continuity of $g$ imply that if
$B$ is not empty, then it is an interval with the left endpoint
zero. It is also clear that $B$ is non-empty \emph{iff} $g(0+)>-1$
which gives the condition $\alpha>-8/\pi$. On the contrary, $B$ is
empty if $\alpha<-8/\pi$ and the borderline case $\alpha=-8/\pi$
was mentioned above.

Let us next focus on the negative part using $\tilde{g}(x):=\cosh
x\pi+\frac{\alpha}{4}\cdot\frac{\sinh x\pi}{x}$ and ask about
$\tilde B:=\{x\in(0,\infty):\: |\tilde g(x)|\leq1\}$. It is easy to check that $\tilde{g}(x)=-1$ iff $\tanh\frac{x\pi}{2}=\frac{4x}{|\alpha|}$ and $\tilde{g}(x)=1$ iff $\coth\frac{x\pi}{2}=\frac{4x}{|\alpha|}$. It implies that there is exactly one $x_1$ such that $\tilde{g}(x_1)=1$, and that the equation $\tilde{g}(x)=-1$ has one solution $x_{-1}$ in the case $\alpha<-8/\pi$ and no solution in the case $\alpha\in[-8/\pi,0)$. Since obviously $0<x_{-1}<x_1$ and $\tilde{g}(0+):=\lim_{x\to0_+}\tilde{g}(x)=1+\alpha\pi/4$,
we infer that $\tilde B$ is a bounded interval. Its closure
contains zero \emph{iff} $\alpha\in[-8/\pi,0)$ because then
$\tilde{g}(0+)\in[-1,1)$. In such a case the
lowest spectral band is the closure of $B\cup\tilde B$, otherwise
it is the closure of $\tilde B$ only. \pfk

\setcounter{equation}{0}
\section{The perturbed system}\label{perturbace}

\subsection{General considerations}

Let us suppose now that the straight chain of the previous section
suffers a bending perturbation as shown in Figure~\ref{ohyb}. We
call the perturbed graph $\Gamma_\vartheta$; it differs from
$\Gamma_0$ by replacing the arc lengths $\pi$ of a fixed ring,
conventionally numbered as zero, by $\pi\pm\vartheta$. The bending
angle $\vartheta$ is supposed to take values from $(0,\pi)$,
regardless of the fact that for $\vartheta\geq2\pi/3$ it is not
possible to consider $\Gamma_\vartheta$ as embedded in the plane
as sketched --- one can certainly realize such a ``bending'' in an
alternative way, for instance, by deforming the selected ring.

The state Hilbert space of the perturbed system is
$L^2(\Gamma_\vartheta)$ and the Hamiltonian is $H_\vartheta$
obtained by a natural modification of $H_0$; our aim is to
determine its spectrum. Since $\Gamma_\vartheta$ has the mirror
symmetry w.r.t. the axis of the zeroth ring passing through the
points $x=\frac12(\pi\pm\vartheta)$, the operator $H_\vartheta$
can be reduced by parity subspaces into a direct sum of an even
part, $H^+$, and odd one, $H^-$; for the sake of simplicity we
drop mostly the subscript $\vartheta$ in the following.

All the components of the wavefunction at energy $k^2$ are linear
combinations of $\e^{\pm\i k x}$. As we have said we use the ring
labelling with zero corresponding to the perturbed one; the mirror
symmetry allows us to study a half of the system only, say, with
non-negative indices. The wavefunction on each ring will be a pair
of functions $\psi_j$ and $\varphi_j$, where $j$ is the circle
index, $\psi_j$ corresponds to the upper semicircle and
$\varphi_j$ to the lower one,
\begin{equation}\label{psij}
\begin{split}
\psi_j(x)&=C_j^+\e^{\i k x}+C_j^-\e^{-\i k x},\quad x\in[0,\pi]\,,\\
\varphi_j(x)&=D_j^+\e^{\i k x}+D_j^-\e^{-\i k x},\quad x\in[0,\pi]
\end{split}
\end{equation}
for $j\in\N$. The situation is different in the case $j=0$, where
the variables run over modified intervals,
\begin{equation}\label{psi0}
\begin{split}
\psi_0(x)&=C_0^+\e^{\i k x}+C_0^-\e^{-\i k x},\quad
x\in\left[\frac{\pi-\vartheta}{2},\pi\right]\\
\varphi_0(x)&=D_0^+\e^{\i k x}+D_0^-\e^{-\i k x},\quad
x\in\left[\frac{\pi+\vartheta}{2},\pi\right]
\end{split}
\end{equation}
There are $\delta$-couplings with the parameter $\alpha$ in the
points of contact, i.e.
\begin{equation}\label{dotyk}
\psi_j(0)=\varphi_j(0)\,\qquad\psi_j(\pi)=\varphi_j(\pi)
\end{equation}
and
\begin{gather}
\psi_j(0)=\psi_{j-1}(\pi)\label{jkjk1}\\
\psi_j'(0)+\varphi_j'(0)-\psi_{j-1}'(\pi)-\varphi_{j-1}'(\pi)
=\alpha\cdot\psi_j(0)\label{jkjk2}
\end{gather}
Substituting~\eqref{psij} into~\eqref{dotyk} we obtain
$$
C_j^+\cdot\sin k\pi=D_j^+\cdot\sin k\pi\quad \textrm{and}\quad
C_j^-\cdot\sin k\pi=D_j^-\cdot\sin k\pi\,,
$$
thus for $k\notin\N_0$ we have $C_j^+=D_j^+$ and $C_j^-=D_j^-$.
The case $k\in\N_0$ can be treated in analogy analogously with the
``straight'' case: it is easy to see that squares of integers are
infinitely degenerate eigenvalues and the eigenfunctions can be
supported by any ring, now with the exception of the zeroth one.
From now on, we suppose $k\notin\N_0$.

Using the coupling conditions ~\eqref{jkjk1} and~\eqref{jkjk2}, we
arrive at a ``transfer matrix'' relation between coefficients of
the neighbouring rings,
\begin{equation}\label{matice}
\left(\begin{array}{c}C_j^+\\C_j^-\end{array}\right)=
\underbrace{\left(\begin{array}{cc} \left(1+\frac{\alpha}{4\i
k}\right)\e^{\i k \pi} & \frac{\alpha}{4\i k}\e^{-\i k \pi}\\
[.3em] -\frac{\alpha}{4\i k}\e^{\i k \pi} &
\left(1-\frac{\alpha}{4\i k}\right)\e^{-\i k \pi}
\end{array}
\right)}_{M}\cdot\left(\begin{array}{c}C_{j-1}^+\\C_{j-1}^-\end{array}\right)\,,
\end{equation}
valid for all $j\geq2$, which yields
\begin{equation}\label{matice2}
\left(\begin{array}{c}C_j^+\\C_j^-\end{array}\right)=
M^{j-1}\cdot\left(\begin{array}{c}C_1^+\\C_1^-\end{array}\right)\,.
\end{equation}
It is clear that the asymptotical behavior of the norms of
$(C_j^+,C_j^-)^T$ is determined by spectral properties of the
matrix $M$. Specifically, let $(C_1^+,C_1^-)^T$ be an eigenvector
of $M$ corresponding to an eigenvalue $\mu$, then $|\mu|<1$ $\:$
($|\mu|>1,\: |\mu|=1$) means that $\left\|(C_j^+,C_j^-)^T\right\|$
decays exponentially with respect to $j\:$ (respectively, it is
exponentially growing, or independent of $j$).

The wavefunction components on the $j$-th ring for both $H^\pm$
(as well as on the (-$j$)-th by the mirror symmetry) are
determined by $C_j^+$ and $C_j^-$, and thus by $(C_1^+,C_1^-)^T$
by virtue of~\eqref{matice2}. If $(C_1^+,C_1^-)^T$ has a
non-vanishing component related to an eigenvalue of $M$ of modulus
larger than one, it determines neither an eigenfunction nor a
generalized eigenfunction of $H^\pm$. On the other hand, if
$(C_1^+,C_1^-)^T$ is an eigenvector, or a linear combination of
eigenvectors, of the matrix $M$ with modulus less than one
(respectively, equal to one), then the coefficients $C_j^\pm$
determine an eigenfunction (respectively, a generalized
eigenfunction) and the corresponding energy $E$ belongs to the
point (respectively, continuous) spectrum of the operator $H^\pm$.
To perform the spectral analysis of $M$, we employ its
characteristic polynomial at energy $k^2$,
\begin{equation}\label{charpol}
\lambda^2- 2\lambda\left(\cos k\pi+\frac{\alpha}{4k}\sin
k\pi\right)+1\,,
\end{equation}
which we have encountered already in the relation \eqref{eik}; it
shows that $M$ has an eigenvalue of modulus less than one
\emph{iff} the discriminant of~\eqref{charpol} is positive, i.e.
$$
\left|\cos k\pi+\frac{\alpha}{4k}\sin k\pi\right|>1\,,
$$
and a pair of complex conjugated eigenvalues of modulus one
\emph{iff} the above quantity is $\leq1$. In the former case the
eigenvalues of $M$ are given by
$$
\lambda_{1\atop 2}=\cos k\pi+\frac{\alpha}{4k}\sin
k\pi\pm\sqrt{\left(\cos k\pi+\frac{\alpha}{4k}\sin
k\pi\right)^2-1}\,,
$$
satisfying $\lambda_2=\lambda_1^{-1}$, hence $\lambda_2<1$ holds
if $\cos k\pi+\frac{\alpha}{4k}\sin k\pi>1$ and $\lambda_1<1$ if
this quantity is $<-1$. Moreover, the corresponding eigenvectors
of $M$ are
$$
v_{1,2}= \left(\begin{array}{c}
\frac{\alpha}{4\i k}\e^{-\i k\pi} \\[.2em]
\lambda_{1,2}-\left(1+\frac{\alpha}{4\i k}\right)\e^{\i k\pi}
\end{array}
\right)\,.
$$

\begin{pozn} \label{ess-inv}
Comparing to~\eqref{sporig} we see that the perturbation does not
affect the spectral bands, and also, that new eigenvalues coming
from the perturbation can appear only in the gaps. These facts are
obvious, of course, from general principles. Using the natural
identification of $L^2(\Gamma_0)$ and $L^2(\Gamma_\vartheta)$ we
see that $H_0$ and $H_\vartheta$ differ by a shift of the point
where a boundary condition is applied, hence their resolvent
difference has a finite rank (in fact, rank two). Consequently,
their essential spectra coincide and each spectral gap of $H_0$
contains at most two eigenvalues of $H_\vartheta$, see
\cite[Sec.~8.3, Cor.~1]{We}.
\end{pozn}

\subsection{Spectrum of $H^+$}

The operator $H^+$ corresponds to the wave functions \emph{even}
w.r.t. the symmetry axis, hence we may consider a half of the
graph with the Neumann conditions at the boundary (i.e., the
points A, B in Figure~\ref{ohyb}),
$$
\psi_0'\left(\frac{\pi-\vartheta}{2}\right)=0\,,
\quad\varphi_0'\left(\frac{\pi+\vartheta}{2}\right)=0\,.
$$
At the contact point of the zeroth and the first ring (denoted by
C) there is a $\delta$-coupling with the parameter $\alpha$,
\begin{gather}
\psi_0(\pi)=\varphi_0(\pi)=\psi_1(0)\\
\psi_1'(0)+\varphi_1'(0)-\psi_0'(\pi)-\varphi_0'(\pi)
=\alpha\cdot\psi_0(\pi)
\end{gather}
Substituting to these conditions from~\eqref{psij}
and~\eqref{psi0} and using the equality
$\varphi_1'(0)=\psi_1'(0)$, we obtain $(C_1^+,C_1^-)^T$ up to a
multiplicative constant,
$$
\left(\begin{array}{c}C_1^+\\C_1^-\end{array}\right)=
\left(\begin{array}{c}
\frac{\cos k\pi+\cos k\vartheta}{\sin k\pi}
+\i\left(1-\frac{\alpha(\cos k\pi+\cos k\vartheta)}{2k\sin k\pi}\right)\\
\frac{\cos k\pi+\cos k\vartheta}{\sin k\pi}
-\i\left(1-\frac{\alpha(\cos k\pi+\cos k\vartheta)}{2k\sin k\pi}\right)
\end{array}
\right)\,.
$$
The right-hand side is well defined except for $\sin k\pi=0$, but
this case has been already excluded from our considerations; we
know that for $k\in\N$ the number $k^2$ is an eigenvalue of
infinite multiplicity.

Following the above discussion $k^2\in\sigma_p(H^+)$ requires that
the vector $(C_1^+,C_1^-)^T$ is an eigenvector of $M$
corresponding to the eigenvalue $\lambda$ of the modulus less than
one. Using the above explicit form of the eigenvectors and solving
the equation
$$
\left|\begin{array}{cc}
\frac{\cos k\pi+\cos k\vartheta}{\sin k\pi}
+\i\left(1-\frac{\alpha(\cos k\pi+\cos k\vartheta)}{2k\sin k\pi}\right)
& \frac{\alpha}{4\i k}\e^{-\i k\pi}\\
\frac{\cos k\pi+\cos k\vartheta}{\sin k\pi}
-\i\left(1-\frac{\alpha(\cos k\pi+\cos k\vartheta)}{2k\sin k\pi}\right)
& \lambda-\left(1+\frac{\alpha}{4\i k}\right)\e^{\i k\pi}
\end{array}
\right|=0
$$
we arrive at the condition
$$
(\cos k\vartheta+\cos k\pi)\cdot
\left(\frac{\alpha}{4k}\sin k\pi
\pm\sqrt{\left(\cos k\pi+\frac{\alpha}{4k}\sin k\pi\right)^2-1}\right)
=\sin^2 k\pi\,,
$$
with the sign given by the sign of $\cos k\pi +\frac{\alpha}{4k}
\sin k\pi$. Since $\sin k\pi\neq0$, the second factor at the
\emph{lhs} is also nonzero and the last equation is equivalent to
\begin{equation}\label{spektrpodm}
\cos k\vartheta=-\cos k\pi +\frac{\sin^2 k\pi}
{\frac{\alpha}{4k}\sin k\pi\pm\sqrt{\left(\cos k\pi
+\frac{\alpha}{4k}\sin k\pi\right)^2-1}}\,;
\end{equation}
for the sake of brevity we denote the expression at the \emph{rhs}
by $f(k)$.

The relation~\eqref{spektrpodm} is our main tool to analyze the
discrete spectrum and we are going to discuss now its solutions.
We start with an auxiliary result noting that, as a consequence of
Theorem~\ref{bands}, the set of positive $k$ for which the
inequality $|\cos k\pi+\frac{\alpha}{4 k}\sin k\pi|\geq1$ is
satisfied is an infinite disjoint union of closed intervals. We
denote them $I_n$ with $n\in\N$ and recall that $n\in I_n$. If
$\alpha>0$ we denote by $I_0$ the interval with the edge at zero
corresponding to the non-negative part of the lowest spectral gap
of $H_0$.
\begin{prop}\label{f}
The function $f$ introduced above maps each $I_n\backslash\{n\}$
into the interval $(-1,1)\cup\{(-1)^{n}\}$. Moreover,
$f(x)=(-1)^n$ holds for $x\in I_n\backslash\{n\}$ \emph{iff}
$|\cos x\pi+\frac{\alpha}{4 x}\sin x\pi|=1$, and $\lim_{x\in
I_n,x\to n}f(x)=(-1)^{n+1}$.
\end{prop}
\pf According to \eqref{spektrpodm}, the function $f$ is
continuous in each interval $I_n\backslash\{n\}$, thus it maps the
interval $I_n\backslash\{n\}$ again to an interval. The claim then
follows from the following easy observations. First,
$f(x)=(-1)^{n}$ \emph{iff} $x$ is the non-integer boundary point
of $I_n$ (if $\alpha<0$ and $|\alpha|$ is sufficiently large, the
left edge of $I_1$ is moved to zero and one checks that
$\lim_{x\to0}f(x)=-1$). Furthermore, for all $x\in
I_n\backslash\{n\}$ we have $f(x)\neq(-1)^{n-1}$, and finally,
$\lim_{x\to n,x\in I_n} f(x)=(-1)^{n-1}$. \pfk

Proposition~\ref{f} guarantees the existence of at least one
solution of~\eqref{spektrpodm} in each interval
$I_n\backslash\{n\}$, except for the case when $\vartheta$
satisfies $\cos n\vartheta=(-1)^{n-1}$, or equivalently, except
for the angles $\vartheta=\frac{n+1-2\ell}{n}\pi$,
$\:\ell=1,\ldots,\left[\frac{n+1}{2}\right]$. Later we will show
that for these angles there is indeed no solution of the equation
\eqref{spektrpodm} in $I_n\backslash\{n\}$, while for the other
angles in $(0,\pi)$ there is exactly one.

In a similar way one can proceed with the negative part of the
spectrum. If $k=\i\kappa$ where $\kappa>0$, the
condition~\eqref{spektrpodm} acquires the form
\begin{equation}\label{spektrpodm-}
\cosh\kappa\vartheta=-\cosh\kappa\pi
-\frac{\sinh^2\kappa\pi}
{\frac{\alpha}{4\kappa}\sinh\kappa\pi\pm\sqrt{\left(\cosh\kappa\pi
+\frac{\alpha}{4\kappa}\sinh\kappa\pi\right)^2-1}}\,,
\end{equation}
where the upper sign in the denominator refers to
$\cosh\kappa\pi+\frac{\alpha}{4\kappa}\sinh\kappa\pi>1$, and the
lower one to $\cosh\kappa\pi +\frac{\alpha}{4\kappa}\sinh\kappa\pi
<-1$. Let us denote the \emph{rhs} of\eqref{spektrpodm-} by
$\tilde{f}(\kappa)$, then we have the following counterpart to
Proposition~\ref{f}.
\begin{prop}\label{tildef}
If $\alpha\geq0$, then $\tilde{f}(\kappa)<-\cosh\kappa\vartheta$
holds for all $\kappa>0$ and $\vartheta\in(0,\pi)$. On the other
hand, for $\alpha<0$ we have \\[.5em]
If $\lim_{\kappa\to0} \left(\cosh\kappa\pi
+\frac{\alpha}{4\kappa}\sinh\kappa\pi\right)<-1$, then there is a
right neighbourhood of zero where $\tilde{f}(x)=-1-C(\alpha)\,
x^2+o(x^2)$ with the constant explicitly given by
$C(\alpha):=\bigg(\frac{1}{2} +\Big(\frac{\alpha\pi}{4}
+\sqrt{\left(\frac{\alpha\pi}{4}\right)^2
+\frac{\alpha\pi}{2}}\Big)^{-1}\bigg)\pi^2$. Moreover,
$\tilde{f}(\kappa)=-1$ holds for $\kappa>0$ \emph{iff}
$\cosh\kappa\pi+\frac{\alpha}{4\kappa}\sinh\kappa\pi=-1$. \\[.5em]
The interval $\left\{\kappa:\: \cosh\kappa\pi
+\frac{\alpha}{4\kappa}\sinh\kappa\pi\geq1\wedge
\kappa\cdot\tanh\kappa\pi<-\alpha/2\right\}$ is mapped by the
function $\tilde{f}$ onto $[1,+\infty)$.\\[.5em]
If $\kappa\,\tanh\kappa\pi>-\alpha/2$, then
$\tilde{f}(\kappa)<-\cosh\kappa\vartheta$ holds for all $\kappa>0$
and $\vartheta\in(0,\pi)$.
\end{prop}

\pf The statement for $\alpha\geq0$ is obvious, assume further
that $\alpha<0$. The first claim follows from the Taylor
expansions of the functions involved in $\tilde{f}$, the last uses
the equality $\cosh^2\kappa-\sinh^2\kappa=1$. The set determined
by the conditions $\cosh\kappa\pi
+\frac{\alpha}{4\kappa}\sinh\kappa\pi\geq1$ and
$\kappa\cdot\tanh\kappa\pi< -\alpha/2$ is obviously an interval
and $\tilde{f}$ is continuous on it. Since $\cosh\kappa\pi
+\frac{\alpha}{4\kappa}\sinh\kappa\pi=1$ implies
$\tilde{f}(\kappa)=1$ and for
$\kappa_0\cdot\tanh\kappa_0\pi=-\alpha/2$ it holds
$\lim_{x\to\kappa_0^-}\tilde{f}=+\infty$, the second claim follows
immediately. Finally, if $\kappa\cdot\tanh\kappa\pi>-\alpha/2$,
then $\cosh\kappa\pi+\frac{\alpha}{4\kappa}\sinh\kappa\pi>1$ and
$\frac{\alpha}{4\kappa}\sinh\kappa\pi\pm\sqrt{\left(\cosh\kappa\pi
+\frac{\alpha}{4\kappa}\sinh\kappa\pi\right)^2-1}>0$, thus
$\tilde{f}(\kappa)<-\cosh\kappa\pi<-\cosh\kappa\vartheta$ holds
for all $\kappa>0$ and $\vartheta\in(0,\pi)$. \pfk

In particular, the first claim concerning $\alpha<0$ together with
the continuity of $\tilde{f}$ implies that if the set
$\left\{\kappa:\:\cosh\kappa\pi+\frac{\alpha}{4\kappa}\sinh\kappa\pi\ge1
\right\}$ is nonempty (and thus an interval), the graph of
$\tilde{f}$ on this set lies below the value -1 touching it
exactly at the endpoints of this interval.

\begin{coro}
If $\alpha\geq0$, then $H^+$ has no negative eigenvalues. On the
other hand, for $\alpha<0$ the operator $H^+$ has at least one
negative eigenvalue which lies under the lowest spectral band and
above the number $-\kappa_0^2$, where $\kappa_0$ is the (unique)
solution of $\kappa\cdot\tanh\kappa\pi=-\alpha/2$.
\end{coro}
\pf The eigenvalues are squares of solutions to the equation
$\cosh\kappa\vartheta =\tilde{f}(\kappa)$. The absence of negative
eigenvalues for $\alpha\geq0$ follows directly from the first
claim in Proposition~\ref{tildef}. The same proposition implies
that there is exactly one interval mapped by $\tilde{f}$ onto
$[1,+\infty)$, hence there is at least one solution of
$\cosh\kappa\vartheta=\tilde{f}(\kappa)$ in this interval. \pfk

\subsection{Spectrum of $H^-$ and a summary}

The operator $H^-$ which corresponds to the odd part of the
wavefunction can be treated in an analogous way. The boundary
conditions on the zero circle are now Dirichlet ones,
$$
\psi_0\left(\frac{\pi-\vartheta}{2}\right)=0\,,\quad
\varphi_0\left(\frac{\pi+\vartheta}{2}\right)=0\,.
$$
One can easily find the spectral condition,
\begin{equation}\label{spektrpodmh-}
-\cos k\vartheta=-\cos k\pi+\frac{\sin^2 k\pi}{\frac{\alpha}{4k}
\sin k\pi\pm\sqrt{\left(\cos k\pi +\frac{\alpha}{4k}\sin
k\pi\right)^2-1}}\,;
\end{equation}
in comparison with~\eqref{spektrpodm} corresponding to $H^+$ there
is a difference in the sign of the cosine on the left-hand side.
Since we already know the behaviour of the right-hand side, cf.
Proposition~\ref{f}, we can infer, similarly as for $H^+$, that
there is at least one solution of~\eqref{spektrpodmh-} in each
interval $I_n$ except for the case when $-\cos
n\vartheta=(-1)^{n-1}$, i.e. when $\vartheta=
\frac{n-2\ell}{n}\pi$, $\:\ell=1,\ldots,\left[\frac{n}{2}\right]$.

Following the analogy with the symmetric case further we can
employ Proposition~\ref{f} to conclude that in each interval $I_n$
there is at least one solution of $-\cos k\vartheta=f(\kappa)$.
The only exception is the interval $I_1$ for $\alpha<0$: for
$|\alpha|$ sufficiently small it holds $-\cos k\vartheta<f(k)$ in
the whole $I_1$; we will comment on this situation in more detail
in the next section devoted to resonances. The negative part of
the point spectrum of $H^-$ is determined by the condition
\begin{equation}\label{spektrpodmhminusneg}
-\cosh\kappa\vartheta=-\cosh\kappa\pi
-\frac{\sinh^2\kappa\pi}{\frac{\alpha}{4\kappa}
\sinh\kappa\pi\pm\sqrt{\left(\cosh\kappa\pi
+\frac{\alpha}{4\kappa}\sinh\kappa\pi\right)^2-1}}\,,
\end{equation}
where we set $k=\i\kappa$ for $\kappa\in\R^+$. It follows from
Proposition~\ref{tildef} that \eqref{spektrpodmhminusneg} has a
solution for negative $\alpha$ only, and it happens if (i) the
positive spectral gap touching zero extends to negative values,
and (ii) the bending angle $\vartheta$ is small enough. In other
words, if there is a number $\kappa_0$ solving
$\cosh\kappa\pi+\frac{\alpha}{4\kappa}\sinh\kappa\pi=-1$, the
energy plot w.r.t. $\vartheta$ obtained as the the implicit
solutions of~(\ref{spektrpodmhminusneg}) is a curve departing from
$(\vartheta,E)=(0,-\kappa_0^2)$; in the next section we will show
that it is analytic and following it one arrives at the point
$(\vartheta,E)=(\pi,1)$.

Let us summarize the discussion of the discrete spectrum. We have
demonstrated that for each of the operators $H^\pm$ there
generally arises at least one eigenvalue in every spectral gap
closure. We have also explained that such an eigenvalue can lapse
into a band edge equal to $n^2$, $n\in\N$, and thus be in fact
absent. The eigenvalues of $H^+$ and $H^-$ may also coincide, in
this case they become a single eigenvalue of multiplicity two. One
can check directly that it happens only if
$$
k\cdot\tan k\pi=\frac{\alpha}{2}\,.
$$

\begin{figure}[h]
\begin{center}
\includegraphics[angle=0,width=0.9\textwidth]{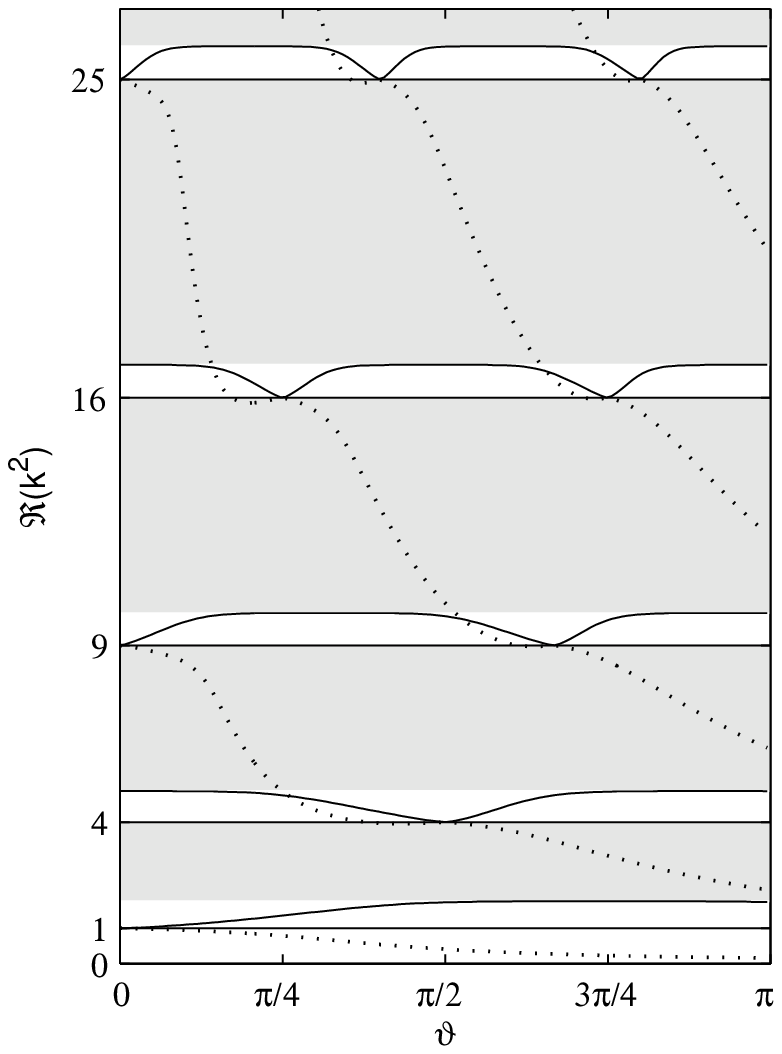}
\caption{The spectrum of $H^+$ as a function of $\vartheta$ for
repulsive coupling, $\alpha=3$. The shaded regions are spectral
bands, the dashed lines show real parts of the resonance pole
positions discussed in Sec.~\ref{reson}.}\label{spec+}
\end{center}
\end{figure}

\begin{figure}[h]
\begin{center}
\includegraphics[angle=0,width=0.9\textwidth]{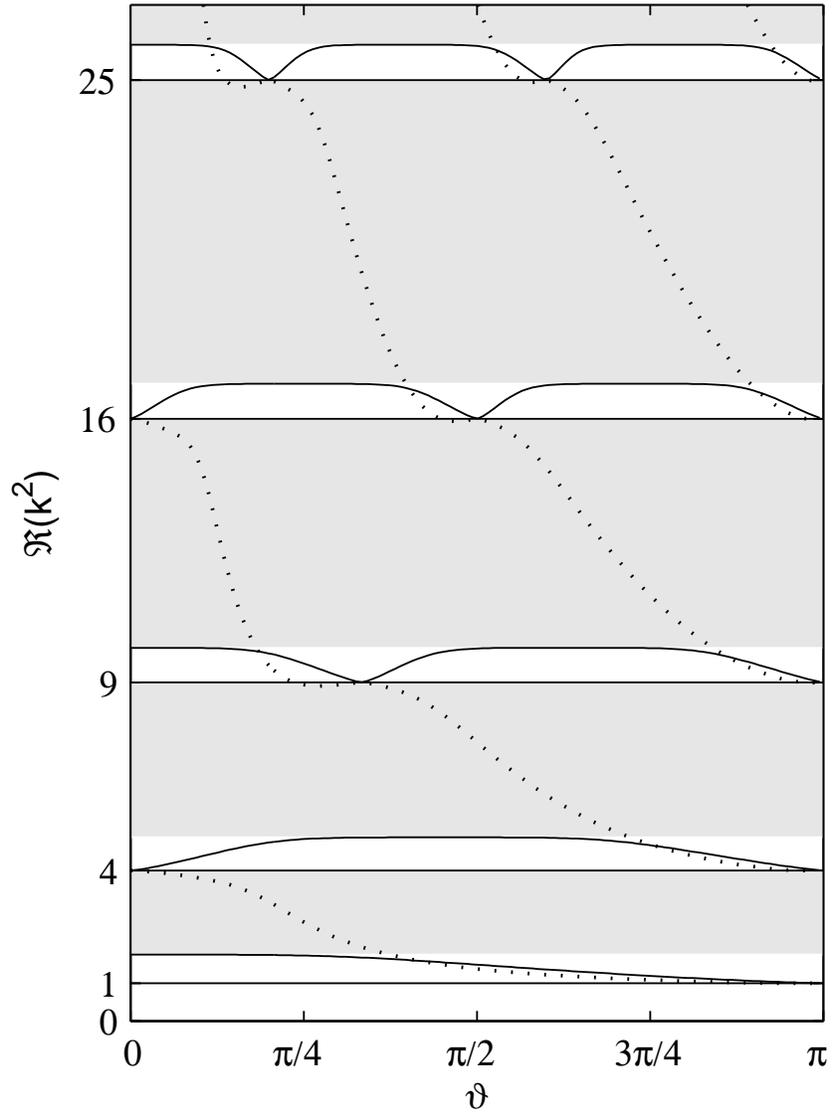}
\caption{The spectrum of $H^-$ in the same setting as in
Fig.~\ref{spec+}}\label{spec-}
\end{center}
\end{figure}

\begin{figure}[h]
\begin{center}
\includegraphics[angle=0,width=0.9\textwidth]{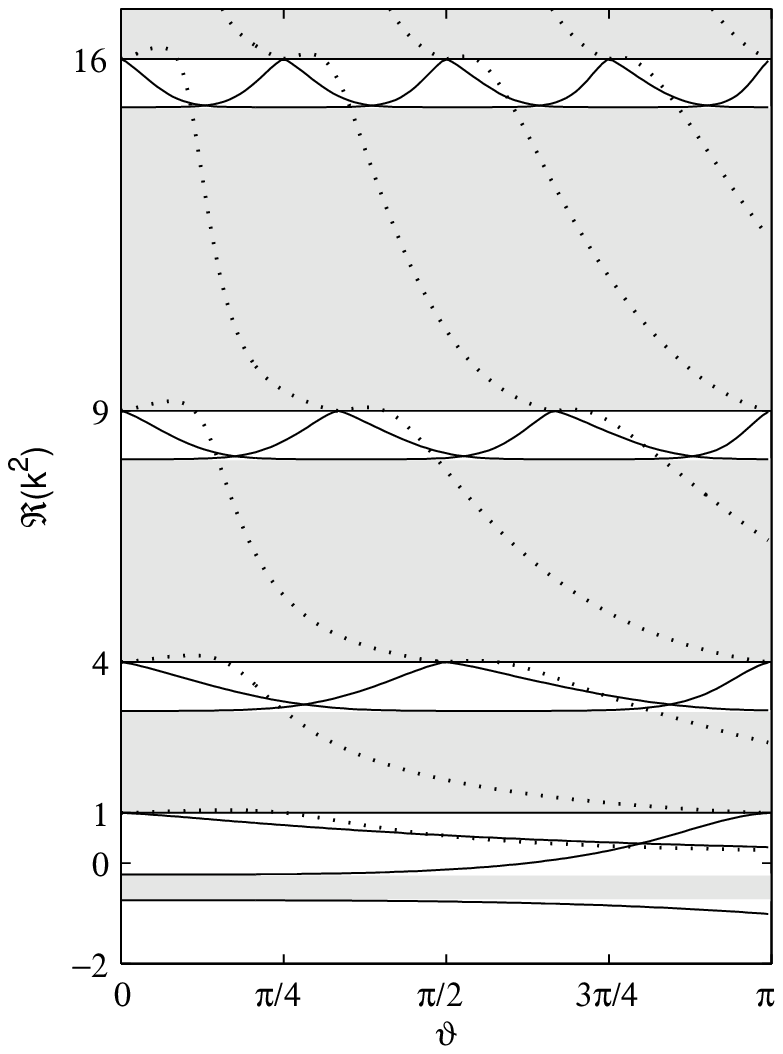}
\caption{The spectrum of $H$ as a function of $\vartheta$ for
attractive coupling, $\alpha=-3$.}\label{attract}
\end{center}
\end{figure}

The study of the resonances, performed in the next section, will
help us to find more precise results concerning the number of
eigenvalues. We will show that there are at \emph{most} two of
them in each spectral gap. However, to make the explanation
clearer, we refer already at this moment to the
Figs.~\ref{spec+}--\ref{attract} illustrating the numerical
solution of the spectral condition for different signs of the
coupling constant, as well as the resonances of the system.

\setcounter{equation}{0}
\section{Resonances and analyticity} \label{reson}

Proceeding further with the discussion we want to learn more about
the angle dependence of the perturbation effects. First we note,
however, that the added eigenvalues are not the only consequence
of the chain bending. One has to investigate all solutions
of~\eqref{rezon+}, not only the real ones which correspond to
$\sigma_p(H^+)$, but also complex solutions describing
\emph{resonances}\footnote{The notion of resonance in the
chain-graph system can be introduced in different, mutually
equivalent, ways similarly as in \cite{EL}.} of $H^\pm$.

\begin{prop}
Given a non-integer $k>0$, the conditions~\eqref{spektrpodm} and
\eqref{spektrpodm-} for $H^\pm$, respectively, are equivalent to
\begin{equation}\label{rezon+}
\frac{\alpha}{2k}(1\pm\cos k\vartheta\cos k\pi)(\pm\cos
k\vartheta+\cos k\pi)=\sin k\pi\cdot(1\pm 2\cos k\vartheta\cos
k\pi+\cos^2 k\vartheta)\,.
\end{equation}
\end{prop}
\pf First we note that changing the square root sign in
denominator of~\eqref{spektrpodm} does not give rise to a real
solution. Indeed, if the sign of the right-hand side
of~\eqref{spektrpodm} is changed, the obtained expression is of
modulus greater than one, hence it cannot be equal to $\cos
k\vartheta$. This further implies that one need not specify the
sign in the denominator of~\eqref{spektrpodm} by the sign of
$\cos k\pi+\frac{\alpha}{4k}\sin k\pi$, and therefore we can
express the square root and subsequently square both sides of the
obtained relation. After simple manipulations, we arrive
at~\eqref{rezon+}; note that for all $k\in\R^+\backslash\N$, the
denominator of~\eqref{spektrpodm} is nonzero. The equivalence
of~\eqref{spektrpodm} and~\eqref{rezon+} for $k\in\C\setminus\N$
is obvious for~\eqref{rezon+} considered with the complex square root, i.e.
without restrictions on the sign in the denominator. The argument
for $H^-$ is analogous. \pfk

Now we are ready to state and prove the analyticity properties.
Since the cases of different symmetries are almost the same, apart
of the position of the points where the analyticity fails, we will
mention the operator $H^+$ only.

\begin{prop}\label{analyticita}
Curves given by the implicit equation~\eqref{rezon+} for $H^+$ are
analytic everywhere except at $(\vartheta,k)
=(\frac{n+1-2\ell}{n}\pi,n)$, where $n\in\N$, $\ell\in\N_0$,
$\ell\leq\left[\frac{n+1}{2}\right]$. Moreover, the real solution
in the $n$-th spectral gap is given by a function
$\vartheta\mapsto k$ which is analytic, except at the points
$\frac{n+1-2\ell}{n}\pi$.
\end{prop}
\pf First we will demonstrate the analyticity of the curves
$\vartheta\mapsto k\in\C$. This is easier done using
equation~\eqref{spektrpodm}; we have to prove that at each point
$(\vartheta,k)$ solving the equation $G(\vartheta,k)=0$ with
$$
G(\vartheta,k)=-\cos k\vartheta-\cos k\pi +\frac{\sin^2
k\pi}{\frac{\alpha}{4k}\sin k\pi \pm\sqrt{\left(\cos
k\pi+\frac{\alpha}{4k}\sin k\pi\right)^2-1}}
$$
the derivative $\frac{\partial G}{\partial\vartheta}$ is nonzero.
We have $\frac{\partial G}{\partial\vartheta} =k\cdot
\sin(k\vartheta)=0$ \emph{iff} $\sin k\vartheta=0$, i.e.
$k\vartheta=m\pi$, $m\in\Z$. This implies $G(\vartheta,k)
=(-1)^{m+1}-\cos k\pi$, and since $G(\vartheta,k)=0$ should be
satisfied, $k$ is an integer of the same parity as $m+1$. For
$k\in\N$, $G$ is not defined and we use~\eqref{rezon+}; it is easy
to check that any solution $(\vartheta,k)$ of~\eqref{rezon+} with
$k\in\N$ corresponds to
$$
\vartheta=\frac{k+1-2\ell}{k}\pi\,,\qquad \ell\in\N,\:
\ell\leq\left[\frac{k+1}{2}\right]\,.
$$
To prove that real solutions are analytic functions, it suffices
to check that, except at the points $(\varphi,k)=(\frac{n+1
-2\ell}{n}\pi,n)$, for each $(\vartheta,k)$ solving
$F(\vartheta,k)=0$ with
 \begin{eqnarray*}
\lefteqn{F(\vartheta,k):=\alpha(1+\cos k\vartheta\cos k\pi) (\cos
k\vartheta+\cos k\pi)} \\ && -2k\sin k\pi\cdot(1+2\cos
k\vartheta\cos k\pi+\cos^2 k\vartheta)
 \end{eqnarray*}
it holds $\frac{\partial F}{\partial k}\neq0$. Computing the
derivative one obtains an expression which can be cast into the
form
\begin{multline*}
2\sin^2 k\pi\cdot(1+2\cos k\vartheta\cos k\pi+\cos^2 k\vartheta)^2+\\
+\alpha\cdot\left[\pi(\cos k\vartheta+\cos k\pi)^4
+\sin^2 k\pi\cdot(\cos k\vartheta+\cos k\pi)^2+\right.\\
\left.+\vartheta\sin^2 k\pi\sin^2 k\vartheta(1+\cos k(\pi-\vartheta))
+(\pi-\vartheta)\sin^2 k\pi\sin^2 k\vartheta(1+\cos k\vartheta\cos k\pi)\right]\,.
\end{multline*}
This is always non-negative, and vanishes \emph{iff}
$$
(\cos k\pi=1\wedge\cos k\vartheta=-1)\vee
(\cos k\pi=-1\wedge\cos k\vartheta=1)\,,
$$
i.e. \emph{iff} $k\in\Z$ and $k\pi=k\vartheta+(2\ell-1)\pi,\:
\ell\in\Z$, proving this the sought claim. \pfk

The resonance dependence on the bending angle $\vartheta$ is again
visualized on Figs.~\ref{spec+}--\ref{attract} where the real
parts are shown; the imaginary parts corresponding to the
situation of Fig.~\ref{spec+} are plotted on Fig.~\ref{imag}
below.

\begin{figure}[h]
\begin{center}
\includegraphics[angle=0,width=0.9\textwidth]{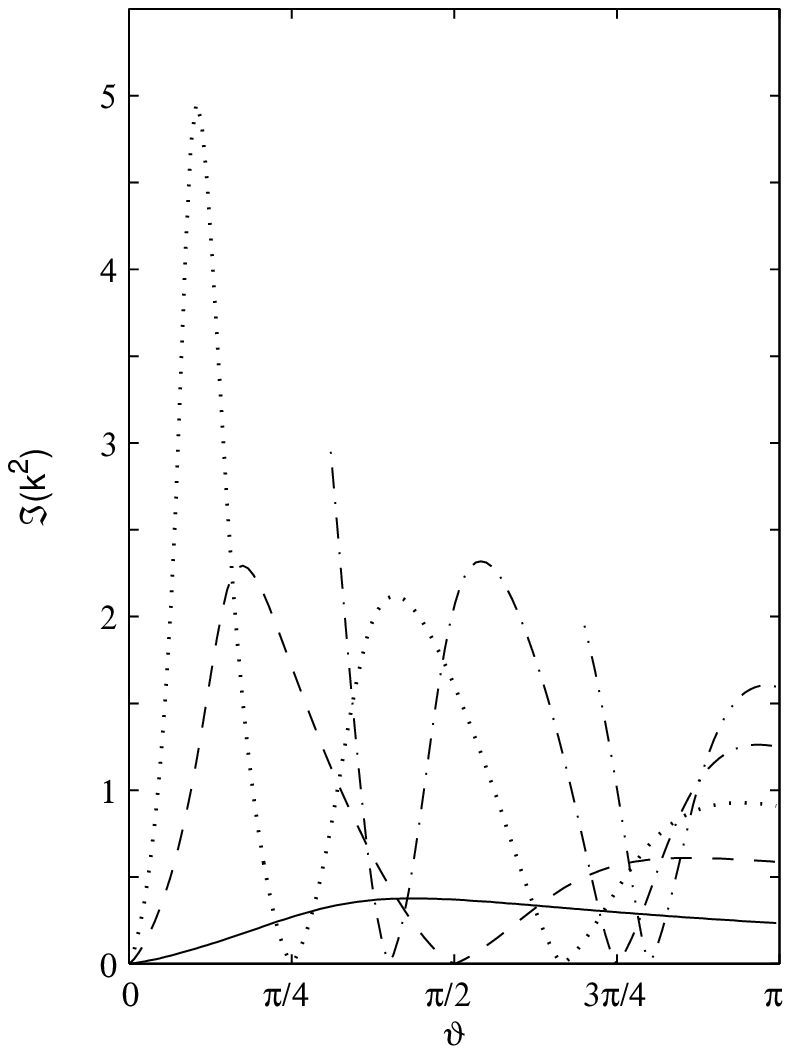}
\caption{The imaginary parts of resonance pole positions in the
same setting as in the previous picture; for the sake of lucidity
only the curves corresponding to $H^+$ are plotted.}\label{imag}
\end{center}
\end{figure}

\setcounter{equation}{0}
\section{More on the angle dependence} \label{angle}

The above results raise naturally the question about the behaviour
of the curves at the singular points $[\vartheta,k]
=[\frac{n+1-2\ell}{n}\pi,n]$ with $n\in\N$, $\ell\in\N$,
$\ell\leq\left[\frac{n+1}{2}\right]$, where they touch the band
edges and where the eigenvalues and resonances may cross. Now we
are going to examine the asymptotic expansion at these points and
to look how many curves ``stem'' from them.

Consider again first the part $H^+$. Let $k_0\in\N$ and
$\vartheta_0:=\frac{n+1-2\ell}{n}\pi$ for some $\ell\in\N$, and
put
$$
k:=k_0+\varepsilon,\qquad \vartheta:=\vartheta_0+\delta\,.
$$
After substituting into~\eqref{rezon+} with the plus signs and
employing Taylor expansions of the $\cos$ and $\sin$ functions we
arrive at the relation
$$
\frac{\alpha}{4} \left(k_0^4\delta^4
+4k_0^3\vartheta_0\delta^3\varepsilon
+6k_0^2\vartheta_0^2\delta^2\varepsilon^2\right)
-k_0\pi^3\varepsilon^3 =\O(\delta\varepsilon^3)
+\O(\varepsilon^4)+\O(\delta^3\varepsilon)\,.
$$
Using the theory of algebroidal functions and Newton polygon, we
find that in the neighbourhood of $(\vartheta_0,k_0)$, the
asymptotical behaviour of solutions is given by the terms of the
order $\delta^4$ and $\varepsilon^3$. In other words, up to
higher-order term we have $\frac{\alpha}{4}k_0^4\delta^4
=k_0\pi^3\varepsilon^3$, and therefore
$$
\left(\frac{\varepsilon\pi}{k_0}\right)^3=\frac{\alpha}{4}\delta^4\,.
$$
Note that $\alpha\in\R$, $k_0>0$, $\delta\in\R$, i.e. only
$\varepsilon$ may be complex here, hence the last equation admits
exactly three types of solutions:
 \begin{itemize}
 \item
$\varepsilon=\sqrt[3]{\frac{\alpha}{4}}\frac{k_0}{\pi}\delta^{4/3}$
(a real solution corresponding to the spectrum)
 \item
$\varepsilon=\e^{\pm\i\frac{2}{3}\pi}\sqrt[3]{\frac{\alpha}{4}}
\frac{k_0}{\pi}\delta^{4/3}$ (imaginary solutions corresponding to
resonances)
\end{itemize}
Let us remark that since~\eqref{rezon+} has a symmetry with
respect to the complex conjugation of $k$, the imaginary solution
come in pairs. This is why we find pairs of curves outside from
the real plane, conventionally just one of them is associated with
a resonance.

Returning to properties of eigenvalues in a fixed spectral gap, we
have so far demonstrated that each real curve describing a
solution of~\eqref{rezon+} is a graph of a function analytic
except at the singular points, cf. Proposition~\ref{analyticita}.
Furthermore, at each singular point only one pair of branches
meets (with respect to the variable $\vartheta$); it follows that
there is exactly one solution in each spectral gap \emph{closure}.
Assuming for definiteness $\alpha>0$ we can say that the complete
graph of solutions of~\eqref{rezon+} has the following structure:
\begin{itemize}
 \item It consists of curves that are analytic and not
intersecting, except at the points $(\vartheta,k)
=(\frac{n+1-2\ell}{n}\pi,n)$, where $n\in\N$, $\ell\in\N$,
$\ell\leq\left[\frac{n+1}{2}\right]$; these are the only
ramification points.
 \item The real curves branches join the
points $(\frac{n+1-2\ell}{n}\pi,n)$ and $(\frac{n+1-2\ell-2}
{n}\pi, n)$, i.e. the consecutive points on the lines $k=n\in\N$.
 \item The curves branches outside the plane $\Im(k)=0$ join the points
\mbox{$(\frac{\ell}{n-\ell}\pi,n-\ell)$} and
$(\frac{\ell+1}{n-\ell-1}\pi,n-\ell-1)$, i.e. the consecutive
points laying on the hyperbolas $(\vartheta+\pi)\cdot
k=n\cdot\pi$, $k\in\R$, $n\in\N$, $n$ odd, cf. Fig.~\ref{imag}.
\end{itemize}
Furthermore, we have seen that the behaviour of eigenvalues in
vicinity of the singular points is the following,
$$
k \approx k_0+ \sqrt[3]{\frac{\alpha}{4}}\frac{k_0}{\pi}\,
|\vartheta-\vartheta_0|^{4/3}\,,
$$
and this is valid for in the particular case $\vartheta_0=0$,
$k_0\in\N$, as well provided the band edge $k_0$ is odd.

However, $H^+$ has an eigenvalue near $\vartheta_0=0$ also in the
gaps adjacent to even numbers. In these cases the curve starts at
the point $(0,k_0)$ for $k_0$ being the solution of $|\cos
k\pi+\frac{\alpha}{4k}\sin k\pi|=1$ in $(n,n+1)$, $\,n$ even. The
asymptotic behaviour of $k$ for $\vartheta$ close to zero is then
different, namely:

\begin{thm} \label{gentle}
Suppose that $n\in\N$ is even and $k_0$ is as described above,
i.e. $k_0^2$ is the right endpoint of the spectral gap adjacent to
$n^2$. Then the behaviour of the solution of~\eqref{rezon+} in the
neighbourhood of $(0,k_0)$ is given by
$$
k=k_0-C_{k_0,\alpha}\cdot\vartheta^4 +\O(\vartheta^5)\,,
$$
where $C_{k_0,\alpha} :=\frac{k_0^2}{8\pi}\cdot
\left(\frac{\alpha}{4}\right)^3 \left(k_0\pi+\sin
k_0\pi\right)^{-1}$.
\end{thm}
\pf The argument is straightforward, it suffices to use Taylor
expansions in~\eqref{rezon+}. \pfk

The analogous asymptotic behaviour applies to $k^2$, the energy
distance of the eigenvalue from the band edge is again
proportional to $\vartheta^4$ in the leading order. Notice that
this is true in any spectral gap, but of course, the error term
depends in general on the gap index.

We refrain from discussing in detail the odd part $H^-$ of the
Hamiltonian. The corresponding results are practically the same,
the only difference is that the roles of the even and odd gaps are
interchanged.

Most of what we have discussed above modifies easily to the case
of attractive coupling with the obvious changes: for $\alpha<0$
the spectral gaps lay now \emph{below} the numbers $n^2$,
$n\in\N$. Of particular interest is the spectral gap adjacent to
the value one, because with the increase of $|\alpha|$ its lower
edge moves towards zero and may become negative for $|\alpha|$
large enough. The even part $H^+$ has similar properties as
before: the eigenvalue curve goes from $(0,1)$ to $(\pi,k_0)$,
where $k_0\in(0,1)$, and there two complex conjugated branches
with $\Re(k)>0$ one of which describes a resonance.

However, the odd part $H^-$ requires a more detailed examination.
We know that there is an eigenvalue curve going to the point
$[\pi,1]$. If the entire spectral gap is above zero, this curve
joins it with $[0,k_0^2]$, where $k_0^2$ is the lower edge of the
gap. On the other hand, if $|\alpha|$ is large enough the
eigenvalue curve starts from $[0,-\kappa_0]$, where $-\kappa_0^2$
is again the lower gap edge; to show that even in this case the
curve joins the points $[0,-\kappa_0]$ and $[\pi,1]$ analytically,
it suffices to prove that the solutions of~\eqref{rezon+} with the
negative sign preserves analyticity when it crosses the line
$k^2=0$.

The spectral condition~\eqref{spektrpodmh-} for $H^-$ is valid for
$k\neq0$. If we put all terms to the left-hand side denoting it as
$\G^-(\vartheta,k)$, i.e.
$$
\G^-(\vartheta,k)=-\cos k\vartheta+\cos k\pi -\frac{\sin^2
k\pi}{\frac{\alpha}{4k} \sin k\pi\pm\sqrt{\left(\cos
k\pi+\frac{\alpha}{4k}\sin k\pi\right)^2-1}}
$$
with the sign in the denominator properly chosen, we have
$\lim_{k\to0}\G^-(\vartheta,k)\,k^{-l}=0\,$ for $l=0,1$ while for
$l=2$ the limit is real-valued and non-vanishing. It follows that
to find the behaviour at the crossing point one has to examine the
function given implicitly by $\tilde{G}(\vartheta,k)=0$, where
$$
\tilde{G}(\vartheta,k)=\left\{\begin{array}{ll}
\frac{\G^-(\vartheta,k)}{k^2} & \textrm{for $k\neq0$}\\ [.3em]
\lim_{k\to0}\frac{\G^-(\vartheta,k)}{k^2} & \textrm{for $k=0$}
\end{array} \right.
$$
This is continuous and it can be easily checked that it has
continuous partial derivatives with respect to $\vartheta$ and $k$
in the neighbourhood of any solution of $\tilde{G}(\vartheta,k)=0$
with $k=0$. In particular, the derivative w.r.t. $\vartheta$
equals $k^{-1} \sin k\vartheta$ for all $k\neq0$, thus at a point
$[\vartheta_0,0]$ solving $\tilde{G}(\vartheta,k)=0$ we have
$$
\frac{\partial\tilde{G}(\vartheta_0,0)}{\partial\vartheta}
=\lim_{k\to0}\frac{\sin k\vartheta_0}{k}=\vartheta_0\neq0\,,
$$
in other words, the solution of $\tilde{G}(\vartheta,k)=0$ is
analytic also at the point $[\vartheta_0,0]$. Needless to say,
this claim which he have checked directly here can be obtained
also by means of the analytic perturbation theory \cite{Ka}.

Finally, note that by Proposition~\ref{analyticita} the solutions
of~\eqref{rezon+} with both the positive and negative signs are
analytic in the whole open halfplane $\Re(k)<0$, and consequently,
no resonances curves can be found there.

\setcounter{equation}{0}
\section{Concluding remarks}

We have reasons to believe that the spectral and resonance
properties due geometric perturbations of the considered type hold
much more generally. In this paper we have decided, however, to
treat the present simple example because it allowed us to find a
rather explicit solution of the problem.

The problem can be viewed from different perspectives. As an
alternative one may interpret the chain graph as a
\emph{decoration} of a simple array-type graph, or if you wish,
the Kronig-Penney model, in the sense of \cite{AI} and \cite{Ku}.
The results of the paper then say that \emph{a local modification}
of the decoration can produce a discrete spectrum in the gaps and
the other effects discussed here.

It is also interesting to draw a parallel between the quantum
graphs  discussed here and \emph{quantum waveguides}, i.e.
Laplacians in tubular domains. Although the nature of the the two
system is very different, they nevertheless share some properties,
in particular, the existence of bound states below the essential
spectrum threshold due to a local bend. This effect is well
studied for Dirichlet quantum waveguides where it is known for a
gentle bend the binding energy is proportional to the fourth power
of the bending angle \cite{DE}, i.e. it has exactly the same
behaviour as described by Theorem~\ref{gentle}.

Bent quantum waveguides with mixed (or Robin) boundary conditions
were also studied \cite{Ji} and it was shown that the effect of
\emph{binding through bending} is present for any repulsive
boundary. In our case an eigenvalue below the lowest band exists
whenever $\alpha\ne 0$ which inspires another look at the
waveguide case. It appears that the argument of \cite{Ji} works
again and proves the existence of curvature-induced bound states
in all cases except the Neumann boundary which is an analogue of
the case $\alpha=0$ here.

\subsection*{Acknowledgment}

The research was supported in part by the Czech Ministry of
Education, Youth and Sports within the project LC06002. One of the
authors (OT) enjoyed a support of the French Government (Bourse du
Gouvernement Francais, Dossier No. 2006 2165) during his stay in
the Centre de Physique Th\'eorique, Marseille, where the major
part of the work has been done.

\end{document}